\documentclass[12pt,preprint]{aastex}
\usepackage{graphicx}
\usepackage{apjfonts}

\newcommand{\apjL}{ApJL}
\newcommand{\sci}{Science}

\begin{document}

\title{{\it Insight}-HXMT Detections of Hard X-ray Tails in Scorpius X-1}

\author{G. Q. Ding\altaffilmark{1}, J. L. Qu\altaffilmark{2}, L. M. Song\altaffilmark{2}, Y. Huang\altaffilmark{2}, S. Zhang\altaffilmark{2}, Q. C. Bu\altaffilmark{3}, M. Y. Ge\altaffilmark{2}, X. B. Li\altaffilmark{2}, L. Tao\altaffilmark{2}, X. Ma\altaffilmark{2}, Y. P. Chen\altaffilmark{2}, Y. L. Tuo\altaffilmark{2}}

\altaffiltext{1}{Xinjiang Astronomical Observatory, Chinese Academy of Sciences, 150, Science 1-Street, Urumqi, Xinjiang 830011, China; dinggq@xao.ac.cn}
\altaffiltext{2}{Key Laboratory of Particle Astrophysics, Institute of High Energy Physics, Chinese Academy of Sciences, Beijing 100049, China}
\altaffiltext{3}{Institut f\"ur Astronomie und Astrophysik, Kepler Center for Astro and Particle Physics, Eberhard Karls Universit\"at, Sand 1, 72076 T\"ubingen, Germany}

\begin{abstract}

Using the observations of the high-energy (HE) detector of the Hard X-ray Modulation Telescope ({\it Insight}-HXMT) for Scorpius X-1 in 2018, we search for hard X-ray tails in the X-ray spectra in $\sim$30-200 keV. The hard X-ray tails are found throughout the Z-track on the hardness-intensity diagram and they harden and fade away from the horizontal branch (HB), through the normal branch (NB), to the flaring branch (FB). Comparing the hard X-ray spectra of {\it Insight}-HXMT between Cyg X-1 and Sco X-1, it is concluded that the hard X-ray spectrum of Cyg X-1 shows high-energy cutoff, implying a hot corona in it, but the high-energy cutoff does not reveal in the hard X-ray spectrum of Sco X-1. Jointly fitting the HE spectrum with the medium-energy and low-energy spectra of Sco X-1 in $\sim$2-200 keV, it is suggested that the upscattering Comptonization of the neutron star (NS) emission photons by the energetic free-falling electrons onto the NS or by the hybrid electrons in the boundary layer between the NS and the accretion disk could be responsible for the hard X-ray tails of Sco X-1 on the HB and NB, but neither of the two mechanisms can be responsible for the hard X-ray tail on the FB. Some possible origins for the peculiar hard X-ray tail of FB are argued.

\end{abstract}

\keywords{Neutron stars (1108); X-ray binary stars (1811); Low-mass X-ray binary stars (939); Non-thermal radiation sources (1119)}

\section{INTRODUCTION}

The hard X-ray emission is usually observed in black hole X-ray binaries (BHXBs) \citep[e.g.,][]{Dove1998,Remillard2006,Montanari2009,Titarchuk2009,Titarchuk2010,Motta2021}, which could be attributed to a hot corona in them \citep[e.g.,][]{Zhang2000,Yao2005,Liu2011,Qiao2012,Yan2020}. In BHXBs, the disk emission X-ray photons illuminating the hot corona are inversely Comptonized by the high-energy electrons in it, resulting in the hard X-ray emission. But, it is uncommon to detect the hard X-ray emission in neutron star low mass X-ray binaries (NS-LMXBs). The so-called hard X-ray tails have been seldom detected in Z sources. However, it is not difficult to detect hard X-ray emission in Atoll sources \citep[e.g.,][]{Piraino1999,Fiocchi2006,Tarana2007,Tarana2011,Raichur2011}. On the other hand, there does not exist any evidence for a hot corona in NS-LMXBs and there is no consensus on the detection of the hard X-ray tails in NS-LMXBs.  

In the past decades, the X-ray satellites with broadband energy ranges have accumulated a great number of observational data for NS-LMXBs, making it possible to search for hard X-ray tails in Z sources. Among these satellites, {\it BeppoSAX} and {\it RXTE}, as well as {\it INTEGRAL}, have contributed a lot on this issue. For example, a hard X-ray power-law (PL) tail having photon index ($\Gamma$) $\sim$2 and contributing $\sim$1.5\% of the source luminosity was significantly  detected in the horizontal branch (HB) of Cyg X-2 from a {\it BeppoSAX} observation \citep{DiSalvo2002}. A similar PL component was also found in the  {\it INTEGRAL} spectrum of GX 5-1 \citep{Paizis2005}. In addition, in the {\it BeppoSAX} spectrum of the third Cyg-like Z source, i.e. GX 340+0, a hard X-ray tail was found to dominate the spectrum above 30 keV  \citep{Lavagetto2004}. As for the first Sco-like Z source, i.e. Scorpius X-1 (Sco X-1),  {\it RXTE} had accumulated 16-year observations,  making it possible to study the properties of the hard X-ray tails of this source in detail. Using the data of {\it RXTE} for Sco X-1, \cite{DAmico2001} and \cite{Ding2021} searched for hard X-ray tails and study their evolution on its hardness-intensity diagram (HID). They found that along the Z-track on the HID, the hard X-ray tail becomes hard and fades away.  The hard X-ray tail was also detected from the {\it INTEGRAL} observations for this source \citep{DiSalvo2006,Revnivtsev2014}. \cite{Farinelli2005,Farinelli2007} detected the transient PL-like hard X-ray tail in one {\it BeppoSAX} observation for the second Sco-like Z source, i.e. GX 17+2 an explained it with bulk-motion Comptonization (BMC) or hybrid Comptonization.  Also using  {\it BeppoSAX} observations,  \cite{DiSalvo2000} searched for the hard X-ray tail of GX 17+2 and investigated its evolution along the Z-track on the HID. They detected the hard X-ray tail with $\Gamma$ $\sim$2.7 in the HB. However, it was not found in the normal branch (NB). Coincidentally, with the {\it RXTE} observations for GX 17+2, \cite{Ding2015} performed the similar investigation and found that the presence of the hard X-ray tail is not confined to a specific branch and it hardens and weakens along the Z-track on the HID, which is similar to the hard tail evolution behavior of Sco X-1 \citep{DAmico2001,Ding2021}. As to the third Sco-like Z source, i.e. GX 349+2, from the {\it BeppoSAX} observation for it,  a hard X-ray tail was also detected in the nonflaring state during its flaring branch (FB)  \citep{DiSalvo2001}.  In addition to the six persistent Z sources, there exists a peculiar Z source, i.e. Cir X-1. Analyzing the {\it BeppoSAX} spectra of Cir X-1, \cite{Iaria2001,Iaria2002} discovered relatively soft hard X-ray tails with $\Gamma$ $\sim$3 at its  periastron and near its apoastron, respectively. Moreover, using the data of {\it RXTE},  \cite{Ding2003,Ding2006b} investigated the hard X-ray tail evolution of Cir X-1 and found that it becomes hard along the Z-track on the HID and it is obviously modulated by the orbit phase.  

It is noted that the two X-ray satellites launched into sky in recent years will enrich the study on the hard X-ray emission of X-ray binaries (XRBs). One is the {\it AstroSat} launched in 2015. The large area xenon proportional counter (LAXPC) of {\it AstroSat} works in the energy range of 3-100 keV  and provides a total of effective area of 8000 $\rm {cm^{2}}$ at 10 keV.  Fitting the LAXPC spectrum of 4U 1538-522, a NS high-mass XRB, \cite{Varun2019} found the cyclotron resonant scattering feature (CRSF) at $\sim$22 keV.  Similarly, with the {\it AstroSat} observations for GRO J2058+42, a transient accreting X-ray pulsar, \cite{Mukerjee2020} detected three CRSFs in the energy range of $\sim$10-45 keV. Another is the Hard X-ray Modulation Telescope ({\it Insight}-HXMT), the first X-ray astronomical satellite of China, launched on June 15, 2017 \citep{Zhang2014,Zhang2020}. {\it Insight}-HXMT consists of three instruments. The high energy (HE) instrument of {\it Insight}-HXMT consists of 18 NaI(TI)/CsI(Na) scintillation detectors with a total of effective area of 5100 $\rm {cm^{2}}$,  working in 20-250 keV \citep{Li2018,Li2020,Li2019}. With the {\it Insight}-HMXT observation for transient accreting pulsar GRO J1008-57, a CRSF at $\sim$90 keV was detected, being the highest energy CRSF \citep{Ge2020}. Another cheering achievement of the HE of {\it Insight}-HXMT is the discovery of quasi-periodic oscillations (QPOs) above $\sim$200 keV in MAXI J1820+070, a BHXB, being the QPO phenomenon at the highest energy range in XRBs \citep{Ma2021}. The two breaking achievements show the excellent ability of the HE instrument of {\it Insight}-HXMT to perform spectral and timing study for XRBs in hard X-ray energy ranges.  

In this work, using the observation data of {\it Insight}-HMXT for Sco X-1 in 2018, we search for hard X-ray tails, compare the hard X-ray spectra between this source and Cyg X-1, and perform investigation on the possible mechanisms for producing the hard X-ray tails in this Z source. We describe our data analyses in Section 2, present our results in Section 3, discuss our results in Section 4, and, finally,  give our conclusions in Section 5.

\section{DATA ANALYSES}

\subsection{Analysis of {{\it Insight}-HXMT} data}

In addition to the HE instrument, {\it Insight}-HXMT includes another two payloads, i.e. the medium energy instrument (ME) and the low energy instrument (LE), which work in 5-30 keV and 1-15 keV, consist of 1728 Si-PIN units and 96 swept charge devices, and have detection areas of 952 $\rm {cm^2}$ and 384 $\rm {cm^2}$, respectively  \citep{Li2018,Li2020,Li2019}.  {\it Insight}-HXMT has accumulated adequate observation data for the Galaxic XRBs and acquired significant achievements \citep[e.g.,][etc.]{Chen2018,Huang2018,Ge2020,Ma2021,You2021}.

The data analysis of any instrument of {\it Insight}-HXMT goes through three stages: calibration $\rightarrow$ screening $\rightarrow$ extracting high level products. At the calibration stage, some 
unusual events, such as spike or trigger events, are removed and the pulse invariant values are calculated from the raw files. Next, during the screening stage,  the screen files, i.e. good event files, are generated with the good time interval (GTI) files which are produced to satisfy some criteria. Finally, at the third stage, the source+background light curves and spectra as well as the background light curves and spectra, together with the spectral response files, are generated. At last, one can get the subtracted-background light curves and spectra. The recommended criteria for producing the GTI files are as follow: (1) the offset angle from the pointing direction <$0.04^\circ$; (2) pointing direction above the earth >$10^\circ$; (3) minimal value of the geomagnetic cutoff rigidity >8; (4) excluding the data in SAA; (5) for LE data, pointing direction above bright the earth >$30^\circ$. The tasks in each stage are finished with some specific tools. For example, the calibration of HE data is executed with the tool HEPICAL, the screening of ME data is implemented with the tool MESCREEN, and extracting the LE spectrum is finished with the tool LESPECGEN, etc.. It is known that the deadtime correction should be considered for analyzing X-ray astronomical data. Thanks to the {\it Insight}-HXMT team, the deadtime correction file for HE data is provided, and that  for ME data is produced in the screening stage, while, the deadtime correction for LE data is unneeded. As for the background, it is worth noting that the {\it Insight}-HXMT team has developed the so-called blind-detector method to estimate the background \citep{Li2018,Li2020,Li2019}.

\subsection{Searching for the hard tails}

\begin{figure}
\vspace{1.2cm}
\centerline{\hbox{
\includegraphics[width=6.5cm,height=8.2cm,angle=-90]{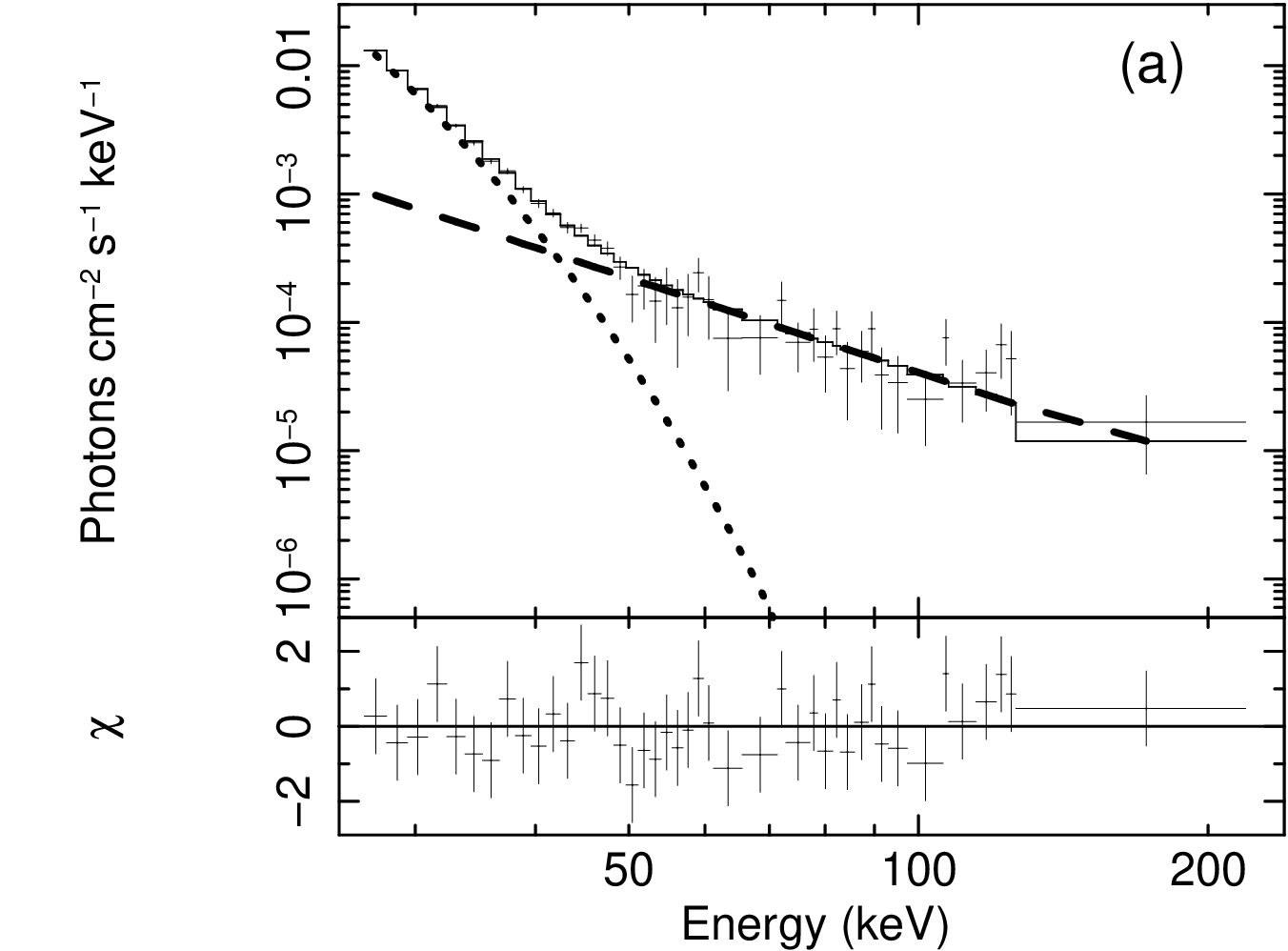}
\hspace{-0.61cm}
\includegraphics[width=6.5cm,height=6.52cm,angle=-90]{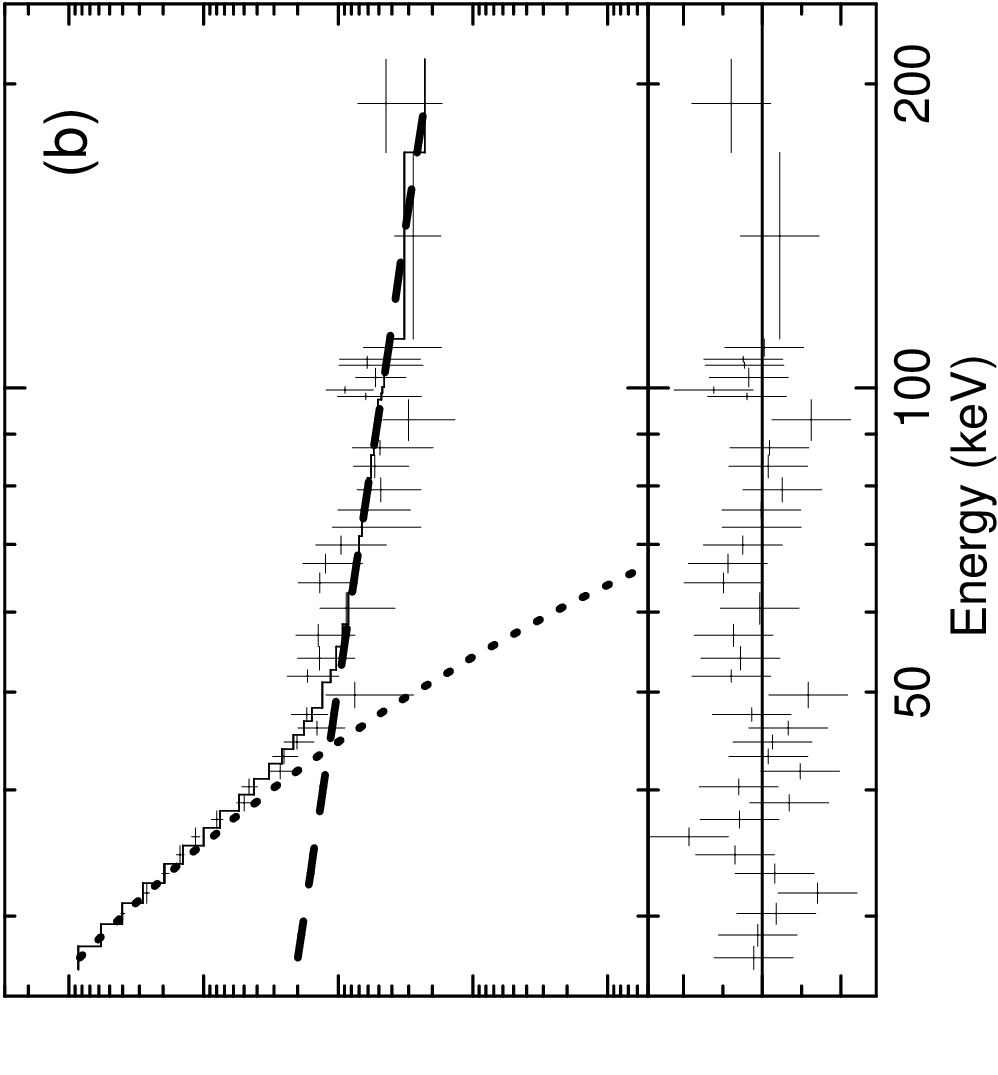}
}}
\vspace{0.25cm}
\caption{The two representive HE spectra of Sco X-1 in which the hard X-ray tail is present. The spectra are fitted by the two-component model consisting of a BREMSS, plus a PL. The individual components are shown, namely,  the BREMSS (the dot line) and the PL (the dash line). Residuals are plotted in units of the standard deviation ($\rm {\sigma}$) with error bars of size one.  Panels (a) and (b) demonstrate the results obtained from OBSIDs P010132801001 and P010132800804, respectively.}\label{fig:hard_tail}
\end{figure}

In this work, we use the {\it Insight}-HXMT data analysis software HXMTSOFT v2.04 to analyze the data. We use the shell script HPIPELINE to generate the light curves and spectra, as well as  the spectral response files, of all the three instrument of each observation at once.. We search for the hard X-ray tails of Sco X-1 from the HE spectra extracted from the observations for this source in 2018. The  {\it Insight}-HXMT data for Sco X-1 in 2018 include 35 observations and, correspondingly, we get 35 HE spectra. In order to improve the signal-to-noise ratios (S/Ns) of the HE spectra, we group the raw channels with S/Ns<1.5, for each new channel resulted from grouping those raw channels to have S/Ns>1.5. It is noted that the raw channels with S/Ns<1.5 are restricted in the energy intervals above $\sim$50-60 keV exclusively. In most cases, after grouping the raw channels above $\sim$50-60 keV, we can only get a few new channels with S/N>1.5, which are distributed around the energy range of  $\sim$50-60 keV, or even cannot get any new channels with S/N>1.5. In these cases,  ignoring the channels below 30 keV and discarding the raw channels with S/Ns<1.5  in the energy bands above $\sim$50-60 keV,  the HE spectra can be fitted well in $\sim$30-60 keV with a thermal emission model such as bremsstrahlung (BREMSS) or COMPTT. Nonetheless,  in a small quantity of cases, after grouping the raw channels above $\sim$50-60 keV, we can get some new channels with S/Ns>1.5, which are distributed in the energy range of $\sim$50-200 keV, as shown in Figure~\ref{fig:hard_tail}. Similarly, ignoring the channels below 30 keV and those raw channels with S/Ns<1.5 above $\sim$200 keV, we perform fittings of these HE spectra in $\sim$30-200 keV. It is found that the obvious high-energy excesses are present in the energy range above $\sim$50-60 keV when these HE spectra are fitted with an one-component model such as BREMSS or COMPTT, etc., and, then, statistically acceptable fittings cannot be obtained, while they are fitted statistically well with a two-component model consisting of a BREMSS or a COMPTT, plus a PL, as demonstrated by Figure~\ref{fig:hard_tail}. Following \cite{DAmico2001} and \cite{Ding2021}, we fit these HE spectra with the two-component model of BREMSS+PL in $\sim$30-200 keV and, meanwhile, perform $F$-test for adding the PL component in the spectral model. The fitting results are listed in Table~\ref{tab:fitting_hard_tail}. Consulting \cite{DAmico2001} and \cite{Ding2021}, we define detection of a hard X-ray tail in the HE spectra satisfying the two needs: (1) the new channels with S/Ns>1.5, resulted from grouping the raw channels, extend over the energy band of  $\sim$50-60 keV considerably; (2) the $F$-test  probability for adding the PL component into the spectral model is less than $\rm {6\times10^{-5}}$.

\section{RESULT}

\begin{figure}[t]
\centerline{
\includegraphics[width=6.cm,height=6.8cm,angle=-90]{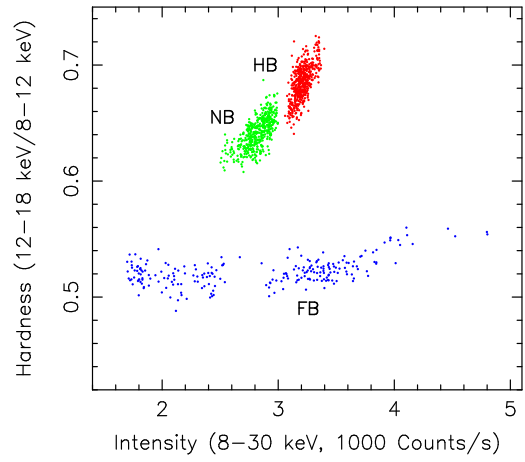}}
\vspace{0.25cm}
\caption{The HID of the eight OBSIDs of Sco X-1 in which hard X-ray tails are detected. The hardness is defined as the ratio of count rate in 12-18 keV to that in 8-12 keV and the intensity is defined as the count rate in 8-30 keV.}\label{Fig:HID_hard_tail}
\end{figure}

\begin{figure}[t]
\centerline{
\includegraphics[width=6.cm,height=7.2cm,angle=-90]{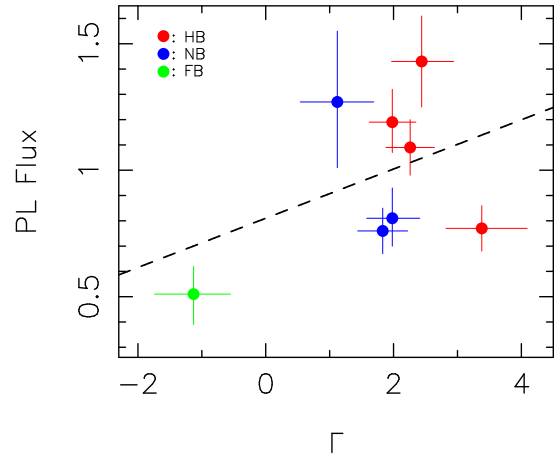}}
\vspace{0.25cm}
\caption{The correlation between the photon index ($\Gamma$) of the hard tail and its flux (in $\sim$30-200 keV, in units of $\rm {10^9\ ergs\ cm^{-1}\ s^{-1}}$).  The red and blue solid circles indicate the results of HB and NB, respectively. The green solid circle represents the result of FB. The dash line is drawn with the least square method.}\label{Fig:Gamma-PL_flux}
\end{figure}

Among the 35 HE spectra of Sco X-1 of {\it Insight}-HXMT, the hard X-ray tails are detected in 8 observations. The HE spectra with detection of a hard X-ray tail are fitted by the two-component model consisting of a BREMSS and a PL. The fitting results, as well as the $F$-test  probability for the PL component to be added into the spectral model, are listed in Table~\ref{tab:fitting_hard_tail}. Two HE spectra with hard tail detection are shown in Figure~\ref{fig:hard_tail}. In order to constrain the positions of the detected hard X-ray tails in the HID, we produce the HID of the 8 observations with hard tail detection using their ME data. We define the hardness as the ratio of count rate in  12-18 keV to that in 8-12 keV, and the intensity as the count rate in 8-30 keV. The produced HID is displayed by Figure~\ref{Fig:HID_hard_tail}. Then, we determine the position of each observation in the HID. As listed in Table~\ref{tab:fitting_hard_tail},  among the 8 detected hard X-ray PL tails, 4 and 3 hard tails are belonged to the HB and NB, respectively, and only one is found in the FB. The values of $\Gamma$ of the HB and NB hard tails are positive, indicating the hard tails in the two branches are relatively hard. However, the $\Gamma$ value of the FB hard tail is negative, being a peculiar hard tail. Such unusual hard tails of Sco X-1 were also found in the high-energy spectra of {\it RXTE} \citep{DAmico2001,Ding2021}. Interestingly, the extraordinary hard tails with negative $\Gamma$ of this source have been always found in the FB. Moreover, it seems that among the Z sources , such peculiar hard tails have been found in Sco X-1 only.  In order to investigate the hard tail evolution on the HID, the correlation between the two parameters of the PL component is demonstrated by Figure~\ref{Fig:Gamma-PL_flux}. Figure~\ref{Fig:Gamma-PL_flux} shows that along the Z-track on the HID, from the HB, via the NB, to the FB, the value of $\Gamma$ decreases and, meanwhile, the PL flux decreases, indicating that the hard tail of Sco X-1 becomes hard and fades away in the sequence HB $\rightarrow$ NB $\rightarrow$ FB, which is consistent with that found by \cite{DAmico2001} and \cite{Ding2021}. Such hard tail evolution behavior was also found in another Z source, i.e. GX 17+2 \citep{Ding2015}.

\section{DISCUSSION}

\subsection{Comparison of hard X-ray spectrum between Cyg X-1 and Sco X-1}

We compare the property of the hard X-ray spectrum  of Cyg X-1 with that of Sco X-1 and try to find some clues for the origin of the nonthermal emission of the latter from the mechanism for producing the hard X-ray emission for the former.

\subsubsection{The hot corona in Cyg X-1}

\begin{figure}
\vspace{1.2cm}
\centerline{\hbox{
\includegraphics[width=6.5cm,height=8.2cm,angle=-90]{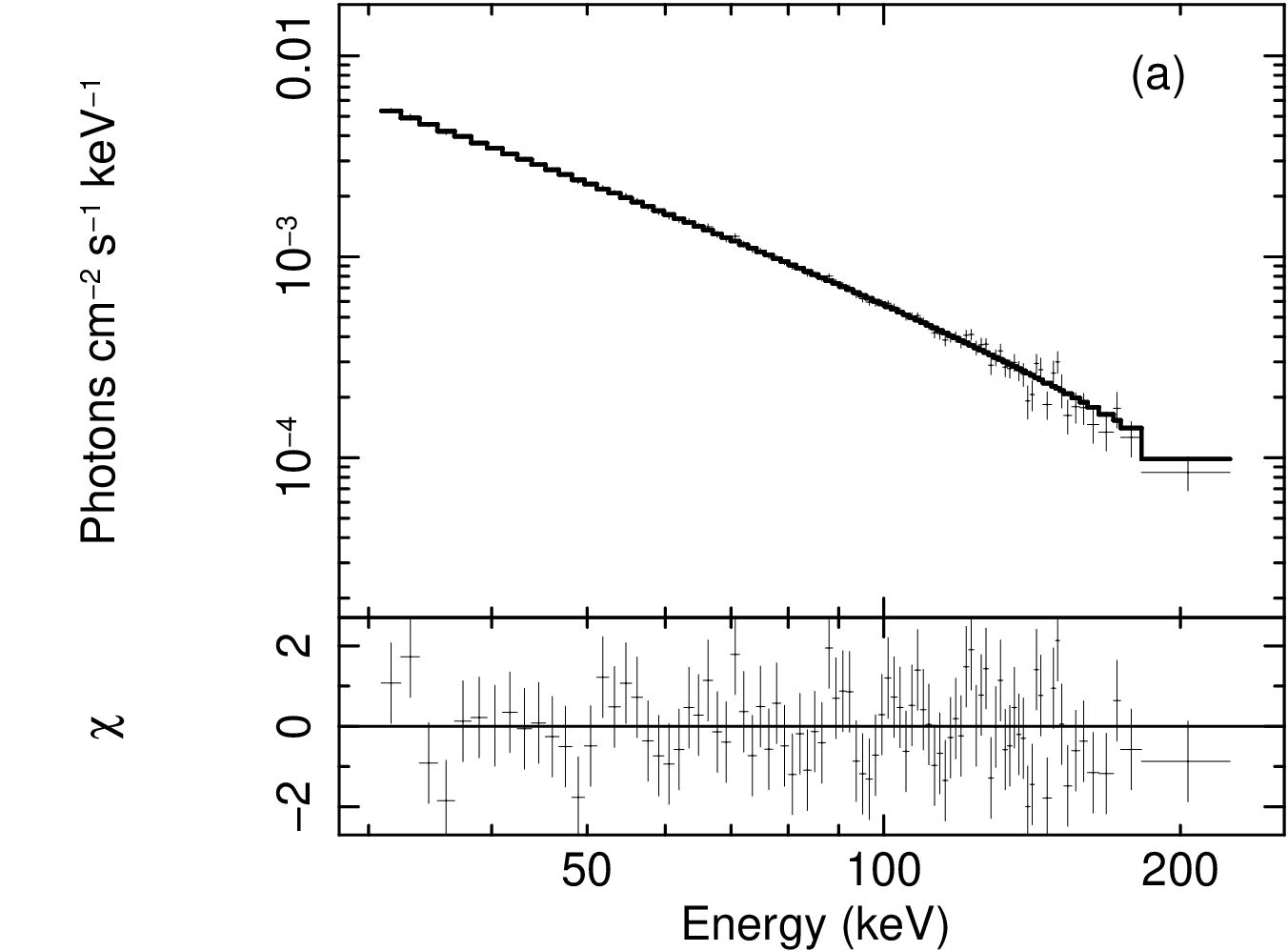}
\hspace{-0.61cm}
\includegraphics[width=6.5cm,height=6.52cm,angle=-90]{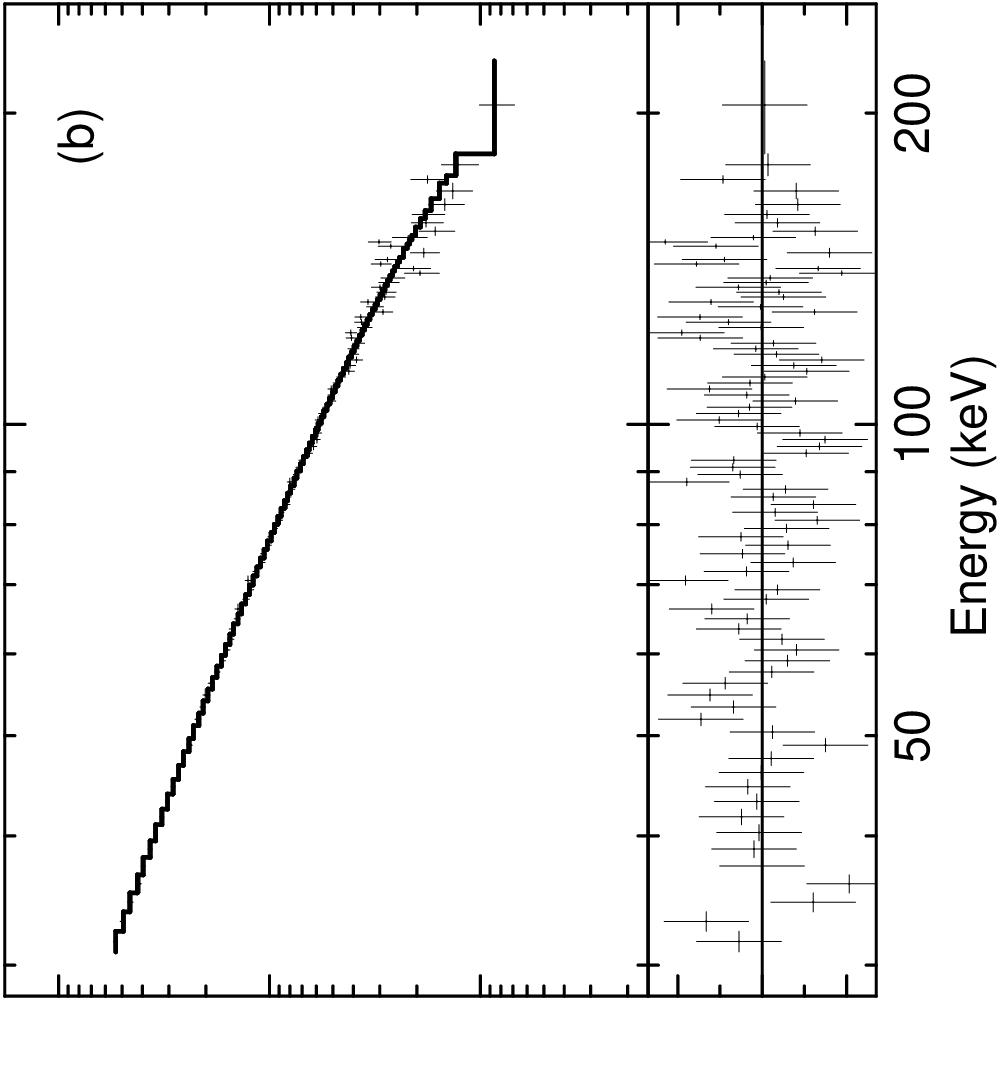}
}}
\vspace{0.25cm}
\caption{The fitting of HE spectrum of Cyg X-1. (a) The spectrum is fitted by the CPL model, with $\chi^2_{\rm {\nu}}$=0.93 for 88 degrees of freedom, $\Gamma=1.53^{+0.06}_{-0.06 }$, and $E_{\rm {cut}}=147^{+0.24}_{-0.28 }$ keV; (b) The spectrum is fitted by the COMPTT model, with $\chi^2_{\rm {\nu}}$=0.90 for 88 degrees of freedom,  $kT_{\rm {e}}=45.4^{+6.5}_{-4.0 }$ keV,  and $\tau=1.16^{+0.12}_{-0.16}$. The parameter errors are derived at the 90\% confidence level $(\Delta\chi^2=2.7)$.}\label{Fig:fitting_HE_CygX-1}
\end{figure}

We extract some HE spectra of {\it Insight}-HXMT of Cyg X-1. We group the raw channels of these HE spectra for each new bin generated from grouping to have S/Ns>5 and, then, fit these grouped spectra in 30-220 keV. It is found that in most cases, the statistically acceptable fittings cannot be obtained when the HE spectra of Cyg X-1 are fitted with a PL, but they can be fitted statistically well by the model of a PL with high-energy exponential cutoff (CUTOFFPL in XSPEC, CPL). On the other hand, in a few cases, although the statistically acceptable fittings can be obtained when the HE  spectra are fitted with a PL or a CPL, the fittings are statistically better fitting with a CPL than fitting with a PL. From our analysis, it is concluded that generally, the hard X-ray spectrum of Cyg X-1 can be fitted statistically well only with a one-component model of CPL, and occasionally, it can be fitted statistically acceptably with a CPL or a PL, in which the fitting is statistically better with a CPL than with a PL. The fitting results with a CPL  of six HE spectra of Cyg X-1 are listed in Table~\ref{tab:HE_Cyg X-1} and a representative fitting is shown by panel (a) of Figure~\ref{Fig:fitting_HE_CygX-1}. The values of $\chi^2_{\rm {\nu}}$ ($\chi^2_{\rm {\nu}}=\chi^2/dof$) indicate that these fittings are statistically well. The presence of high-energy cutoff indicates that the hard X-ray emission of Cyg X-1 is originated from thermal emission. Therefore, we fit the six spectra with the thermal Comptonization models in XSPEC, e.g. COMPTT \citep{Titarchuk1994} or COMPST \citep{Sunyaev1980}, etc. It is found that each of the six HE spectra can be fitted by these models well and the fitting results with the COMPTT model are listed in Table~\ref{tab:HE_Cyg X-1}. Panel (b) of Figure~\ref{Fig:fitting_HE_CygX-1} shows a fitting of compTT. It is noted that the cutoff energy ($E_{\rm {cut}}$) and the electron temperature ($kT_{\rm {e}}$) span a range of $\sim$150-200 keV and a range of $\sim$45-85 keV, respectively. Interestingly, $E_{\rm {cut}}$ is related to $kT_{\rm {e}}$ in this way $E_{\rm {cut}}\approx2.2-3.5\ kT_{\rm {e}}$, approximating the anticipated relation $E_{\rm {cut}}=2-3\ kT_{\rm {e}}$ \citep{Petrucci2001,Yan2020}. The fact that the temperature of the electrons for Comptonization is so high implies that there exists a hot source in Cyg X-1 to supply these high-temperature electrons. Generally, such a hot source is assumed to be a hot corona in XRBs. Moreover, the hot corona in XRBs should be apart from the accretion disk and it might be above the cool disk. Otherwise, the disk will be evaporated by the hot corona and it will disappear. Such a hot corona in BHXBs could be responsible for the frequently observed hard X-ray emission in them \citep[e.g.,][]{Zhang2000,Yao2005,Liu2011,Qiao2012,Yan2020}. The soft X-ray photons from the disk are inversely Comptonized by the high-energy electrons in the hot corona to produce hard X-ray emission.

In accreting black hole systems, forming a corona above the accretion disk could be due to the large-scale magnetic field in them. In these systems, along the open magnetic filed lines, the accreting matter will be hurled to a position above the disk \citep{Blandford1977,Blandford1982}. It is possible for the hurled plasma to accumulate and, then,  constitute a corona at the location, where the radiation pressure and magnetic pressure balance the gravitation. The corona located above the disk might be heated and become into a hot corona through the magnetic buoyancy mechanism, in which the magnetic field lines float above the disk and heat the corona through magnetic reconnection \citep{Galeev1979,Liu2002,Liu2003}.

\subsubsection{A possible warm corona in Sco X-1}

As analyzed above, the HE spectrum of Cyg X-1 shows high-energy cutoff, which indicates a hot corona. However, the HE spectrum of Sco X-1 does not present high-energy cutoff, which rules out  a hot corona in it. Nevertheless,  it is worthy of investigating whether or not there exists a warm corona in it. Therefore, we fit the low energy spectrum of Sco X-1. Generally, there are two types of models to fit the low-energy spectrum of NS-LMXBs, i.e. the so-called Eastern model and Western model.  Each of the two models includes two continua. The Eastern model consists of a blackbody (BB) to account for the emission from the NS surface or a Comptonized BB \citep[COMPBB,][]{Nishimura1986} to describe the Comptonization emission around the NS, plus a multicolored disk (MCD) to be responsible for the emission  from the disk \citep{Mitsuda1984,Mitsuda1989}.  Therefore, this model can be written in two forms, i.e. BB+MCD and COMPBB+MCD. In the Western model, the NS emission is described by a BB and, however, the disk emission is interpreted as an unsaturated Comptonized continuum approximated with a CPL or a Sunyaev-Titarchuk Comptonization \citep[COMPST,][]{Sunyaev1980} \citep{White1985,White1986}. Thus, the Western model can also be expressed in two forms, i.e. BB+CPL and BB+COMPST. Taking into account the interstellar absorption by fixing the hydrogen column density $N_{\rm {H}}$  at $\rm {0.3\times 10^{22}\ atom\ cm^{-2}}$ \citep{Church2012} and the iron line in the spectrum, we use the Eastern and Western models to fit the LE+ME spectra of Sco X-1 in 2-35 keV. Our analysis suggests that the LE+ME spectrum cannot be fitted by any form of the Eastern model, but it can be fitted by any of the two forms of the Western model statistically and physically well. Moreover,  if the COMPST is replaced with a COMPLS \citep{Lamb1979} or a  COMPTT in the form of BB+COMPST, the spectrum can also be fitted well. Nevertheless, the fitting is statistically better with the BB+CPL form than with any other forms, because among these fittings,  the lest value of $\chi^2_ {\rm {\nu}}$ is obtained from the fitting with the BB+CPL form. A representative  fitting with the BB+CPL form is demonstrated by Figure~\ref{Fig:fitting_LE+ME_spectrum}. The fitting results are consistent with those obtained through from fitting the {\it EXOSAT} or {\it RXTE} spectra with the same model for the same source \citep{White1985,Church2012}. The BB temperature ($kT_ {\rm {bb}}$) is $\sim$3 keV, indicating that the BB emission is indeed from the surface of the neutron star (NS), because in NS-LMXBs, generally, the temperature of the NS surface emission is above $\sim$2 keV, while the temperature of the disk emission is below $\sim$1.5 keV \citep[e.g.,][etc.]{Mitsuda1984,Mitsuda1989,Ding2006b,Ding2011,Church2012}. In the Western model, \cite{White1986} proposed that the CPL describes the unsaturated Comptonization and the Comptonization process takes place in the inner disk region.  However, \cite{Church1995,Church2004} explained the CPL component in this way that the soft X-ray photons from the disk are Comptonized by the electrons from a corona on the disk. They called the corona on the disk as an accretion disk corona (ADC) and developed the Western model into the so-called ADC model. It is noted that the $E_{\rm {cut}}$ of CPL is $\sim$4.5 keV, so the temperature of the electrons for Comptonization ($kT_{\rm {e}}$) is only several keV. Therefore, if there does exist a corona in Sco X-1, the corona could be a warm corona rather than a hot corona as that in Cyg X-1. Sequentially, it is unlikely for such a warm corona in Sco X-1 to provide so energetic electrons for Comptonization to produce hard X-ray emission. 

As demonstrated by Fig. 1 of \cite{Lamb1973} or Fig. 1 of \cite{Ghosh1979}, in NS-LMXBs, owing to the small-scale magnetic field, the magnetic field lines form an enclosed structure around the NS and, thus, the plasma can only move in the space near the NS, leading to that the accreting matter cannot be expelled far away. Therefore, in NS-LMXBs, if a corona is formed, it must reside near the accretion disk. The corona should be a warm one, as supported by our analysis above. Otherwise, the disk will be evaporated and disappeared.  

\subsection{The origin of the hard X-ray emission of Sco X-1}
\subsubsection{The thermal component in the hard X-ray spectrum of Sco X-1}

As shown in Figure~\ref{fig:hard_tail}, the HE spectrum of Sco X-1 consists of a PL component, plus a thermal component. The thermal component dominates the emission in $\sim$30-50 keV and it certainly originates from the thermal emission in the system. There are two thermal emission sources in NS-LMXBs, i.e. the NS and the accretion disk. In order to distinguish between the two sources, we replace the BREMSS with a BB or a MCD in the BREMSS+PL model to fit the HE spectrum of Sco X-1. Our fitting suggests that the spectrum can be fitted statistically well either by the BB+PL model, as shown with panel (a) of Figure~\ref{Fig:fitting_HE_ScoX-1}, or by the MCD+PL model. The PL parameters obtained from fitting with the two models are similar to those obtained from the fitting with the BREMSS+PL model. The BB temperature ($kT_{\rm {bb}}$) obtained through fitting with the BB+PL model and the inner disk temperature ($kT_{\rm {in}}$) gotten by fitting with the MCD+PL model are $\sim$3 keV and  $\sim$3.5 keV, respectively. The $kT_{\rm {bb}}$ is significant, but the $kT_ {\rm {in}}$ is meaningless, because, generally, in NS-LMXBs,  the BB temperature of the NS surface emission is around or above $\sim$2 keV, but the inner disk temperature  is below $\sim$1.5 keV \citep[e.g.,][etc.]{Mitsuda1984,Mitsuda1989,Ding2006a,Ding2011,Church2012}. Furthermore, we replace the BREMSS with the COMPTT in the BREMSS+PL model and use the COMPTT+PL model to fit the HE spectrum and get statistically acceptable fitting too, which is demonstrated by panel (b) of  Figure~\ref{Fig:fitting_HE_ScoX-1}. From this fitting, the obtained seed photon temperature ($kT_ {\rm {0}}$), the plasma temperature ($kT$), and the plasma optical depth are $\sim$2.2 keV, $\sim$3.8 keV, and 2.5, respectively, which could be resulted from the Comptonization emission that the photons of the BB emission of the NS surface are thermally Comptonized in the boundary layer between the NS and the disk \citep{Barret2000}. Therefore, it is concluded that the thermal component in the hard X-ray spectrum of Sco X-1 comes from the NS surface emission or the Comptonization emission with the NS surface emission photons as the seed ones. In other words, the thermal component in the HE spectrum is related to the NS emission, instead of the disk emission.

\subsubsection{The nonthermal component in the hard X-ray spectrum of Sco X-1}

\begin{figure}[t]
\centerline{
\includegraphics[width=6.5cm,height=8.2cm,angle=-90]{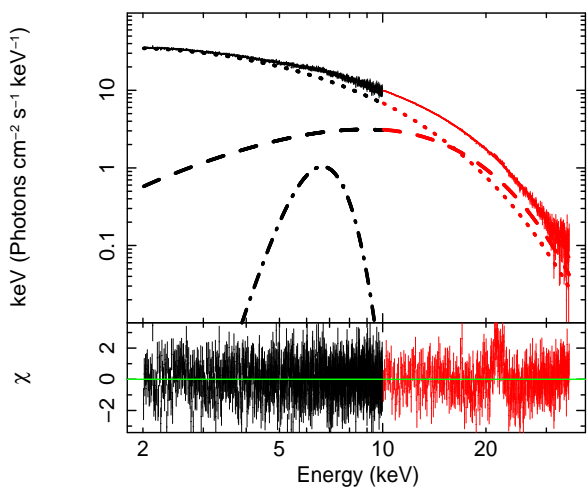}}
\vspace{0.25cm}
\caption{A representative fitting of the LE+ME spectrum of Sco X-1 with the BB+CPL form, plus the line component. The fitting is performed in 2-35 keV. The black and red colours represent the results obtained from the LE and ME instruments of {\it Insight}-HXMT, respectively. Individual components are shown, namely, BB (dash line), LINE (dash-dot line, gaussian line profile), and CPL (dot line). The $\chi^2_{\rm  {\nu}}$ is 1.04 for 1361 degrees of freedom. The fitting parameters: $kT_{\rm {bb}}=3.11^{+0.02}_{-0.03 }$ keV for the BB; $E_{\rm  {Fe}}=6.54^{+0.10}_{-0.10 }$ keV and $\sigma_{\rm  {Fe}}=0.93^{+0.18}_{-0.15 }$ keV for the LINE;  $\Gamma=1.00^{+0.03}_{-0.04 }$ and $E_{\rm {cut}}=4.55^{+0.24}_{-0.28 }$ keV for the CPL. The parameter errors are inferred at the 90\% confidence level $(\Delta\chi^2=2.7)$.}\label{Fig:fitting_LE+ME_spectrum}
\end{figure}

\begin{figure}
\vspace{1.2cm}
\centerline{\hbox{
\includegraphics[width=6.5cm,height=8.2cm,angle=-90]{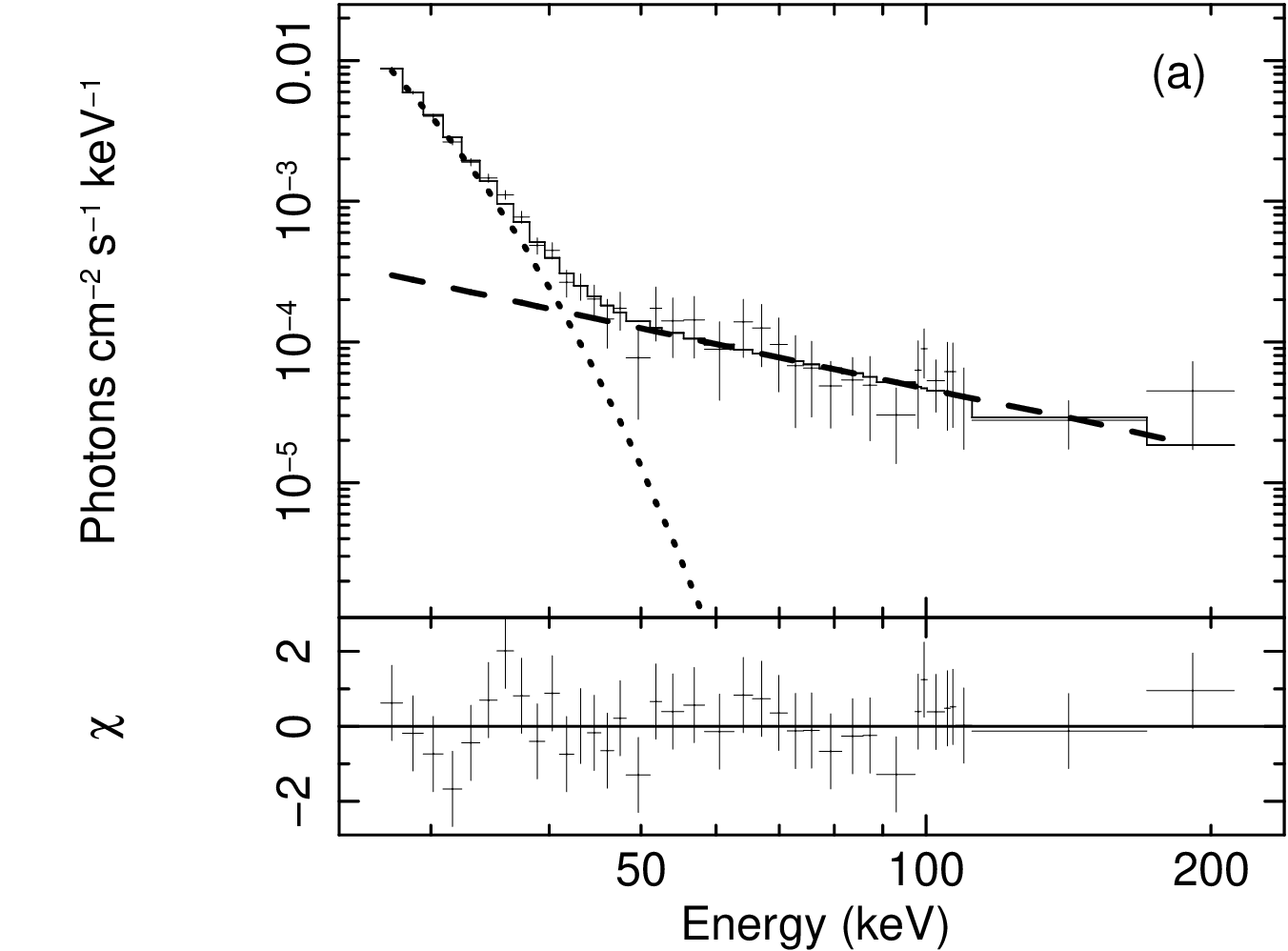}
\hspace{-0.61cm}
\includegraphics[width=6.5cm,height=6.52cm,angle=-90]{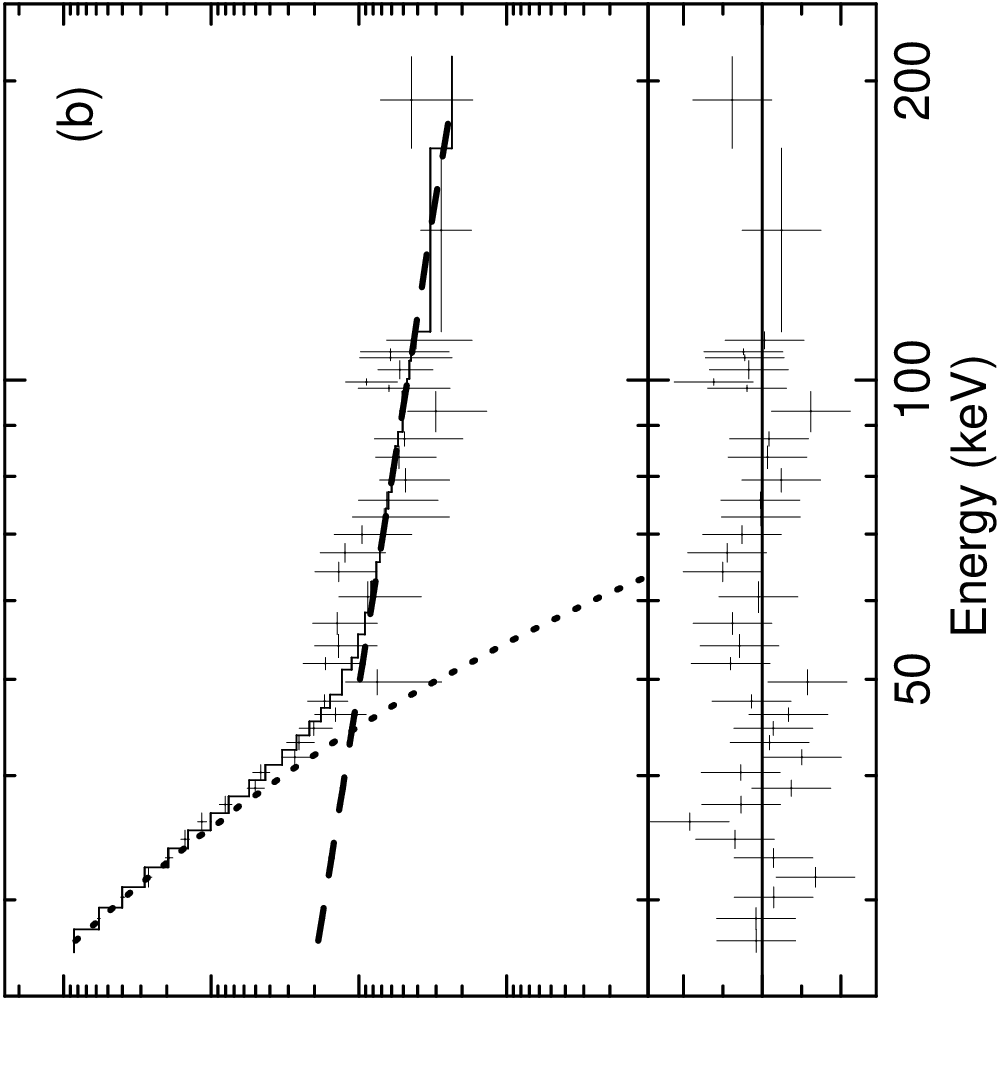}
}}
\vspace{0.25cm}
\caption{The fitting of HE spectrum of Sco X-1. (a) The spectrum is fitted by the model consisting of a BB, plus a PL. The $\chi^2_{\rm {\nu}}$ is 0.63 for 33 degrees of freedom, with $kT_{\rm {BB}}=2.95^{+0.12}_{-0.12 }$ keV for the BB and $\Gamma=1.43^{+0.54}_{-0.53 }$ for the PL. The individual components are shown, namely, a BB (dot line) and a PL (dash line). (b) The spectrum is fitted by the two-component model of COMPTT+PL. The $\chi^2_{\rm {\nu}}$ is 0.58 for 32 degrees of freedom. The individual components are displayed, i.e. a COMPTT (dot line) and a PL (dash line). The seed photon temperature ($kT_{\rm {0}}$) and the electron temperature ($kT_{\rm {e}} $) are 2.15 keV (fixed) and $3.81^{+2.13}_{-0.89}$ keV, respectively, for the COMPTT. The photon index ($\Gamma$) is $1.07^{+0.67}_{-0.74}$ for the PL. The parameter errors are inferred at the 90\% confidence level $(\Delta\chi^2=2.7$.}\label{Fig:fitting_HE_ScoX-1}
\end{figure}

\begin{figure}
\vspace{1.2cm}
\centerline{\hbox{
\includegraphics[width=6.5cm,height=8.2cm,angle=-90]{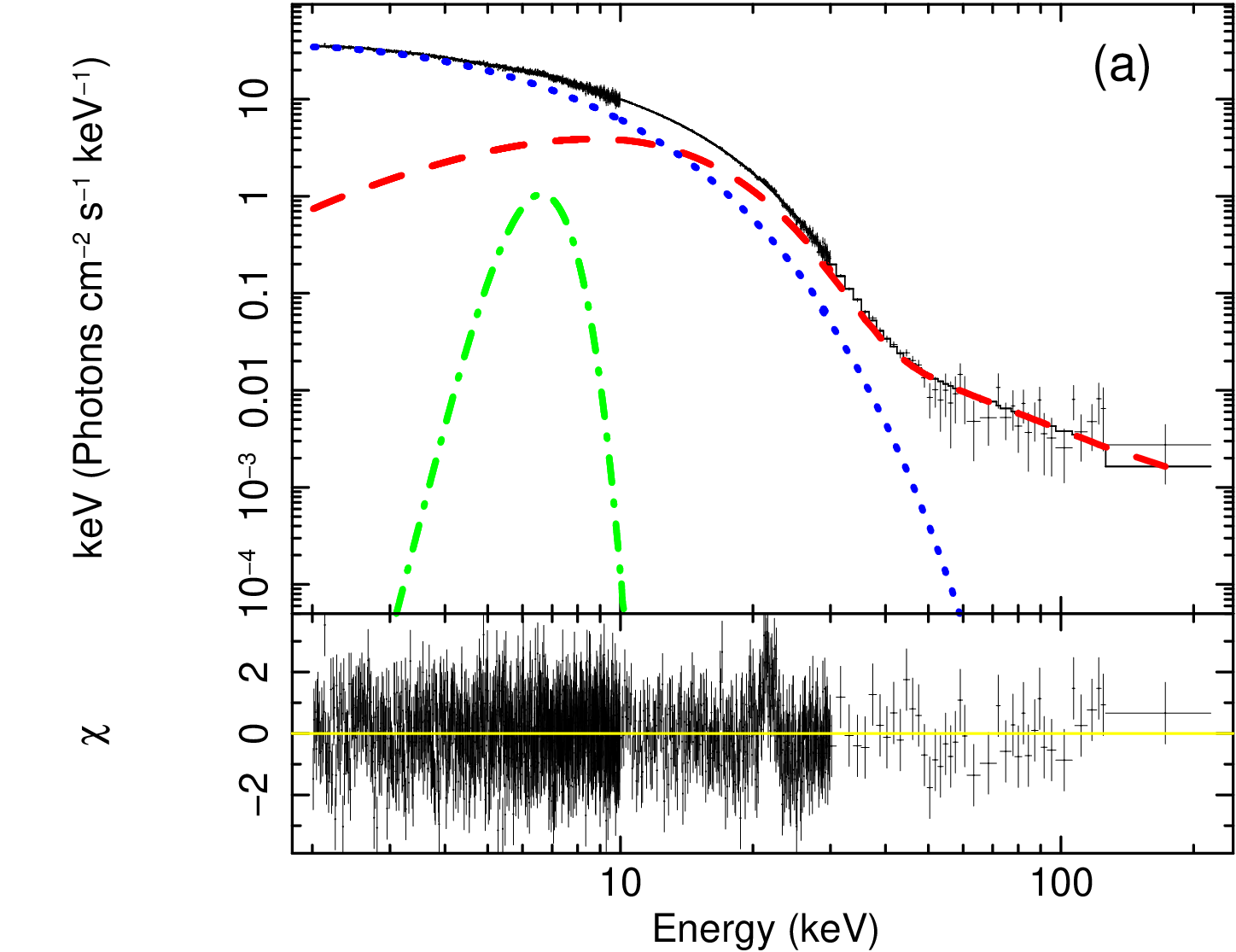}
\hspace{-0.61cm}
\includegraphics[width=6.5cm,height=6.52cm,angle=-90]{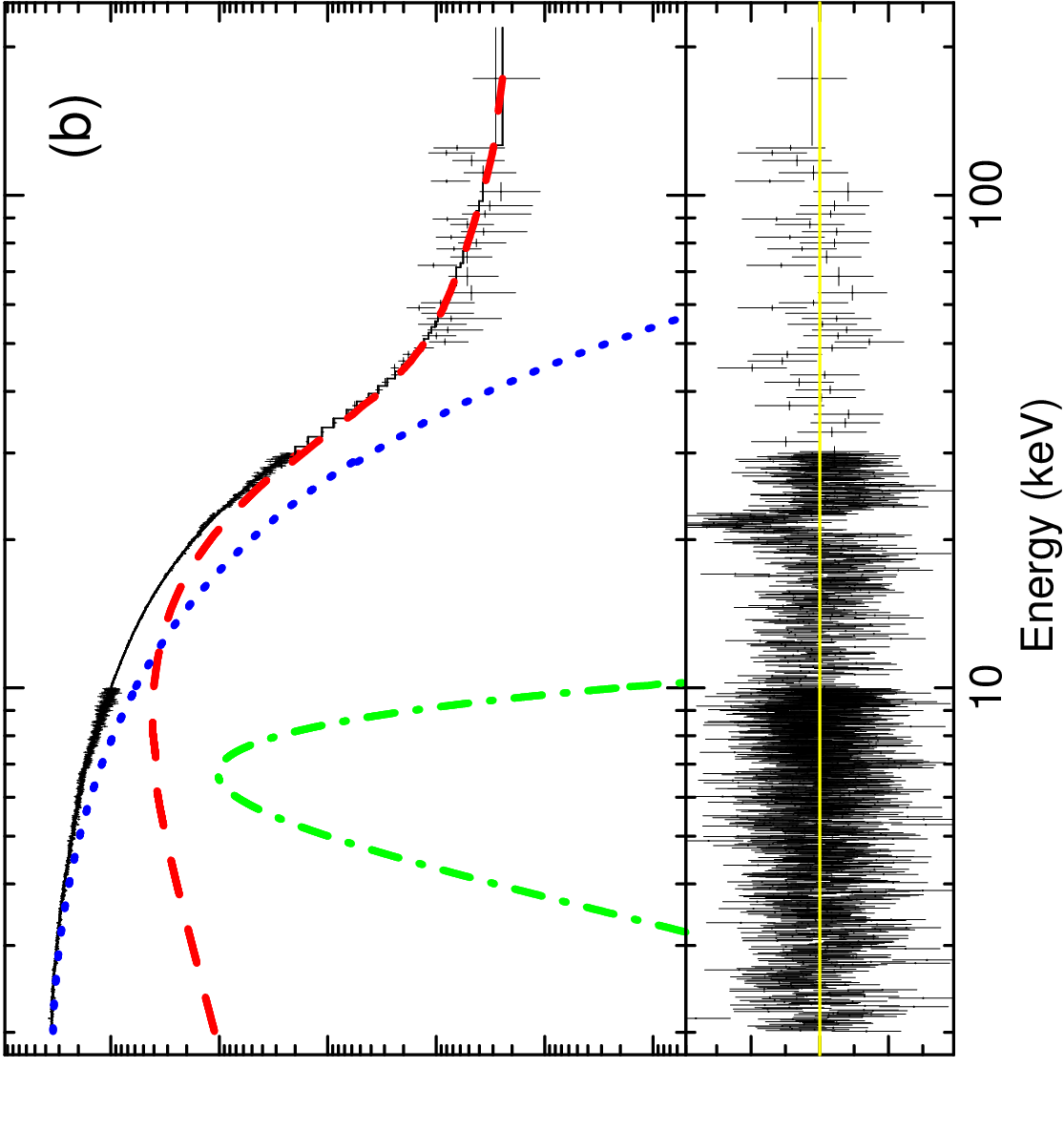}
}}
\vspace{0.25cm}
\caption{The fitting of the LE+ME+HE spectrum of Sco X-1 in $\sim$2-200 keV, in which a hard X-ray tail is detected. (a) The spectrum is fitted by the BMC+LINE+CPL model. The individual components are shown, namely, a BMC (red dash line), a LINE (green dash-dot line), and a CPL (blue dot line). The $\chi^2_{\rm  {\nu}}$ is 1.06 for 1315 degrees of freedom. The fitting parameters: $kT_{\rm {bb}} $=$3.04^{+0.02}_{-0.02}$ keV, $\alpha$=$1.76^{+0.36}_{-0.32}$,  $logA$=$-1.48^{+0.14}_{-0.13}$, and  $N$=$1.04^{+0.11}_{-0.11}$ for the BMC; $E_{\rm {Fe}} $=$6.53^{+0.09}_{-0.09}$ keV, $\sigma_{\rm {Fe}} $=0.8 keV (fixed), and $N$=$0.32^{+0.03}_{-0.03}$ for the LINE; $\Gamma$=$0.96^{+0.04}_{-0.04}$,  $E_{\rm {cut}}$=$4.18^{+0.27}_{-0.26}$ keV, and $N$=$61.89^{+0.56}_{-0.55}$ for the CPL. (b) The spectrum is fitted by the EQPAIR+LINE+CPL model. The individual components are shown, namely, a EQPAIR (red dash line), a LINE (green dash-dot line), and a CPL (blue dot line). The $\chi^2_{\rm {\nu }}$ is 1.08 for 1314 degrees of freedom. The main fitting parameters for the EQPAIR: $l_{\rm {h}}/l_{\rm {s}}$=$\rm {0.15\pm 0.03}$, $l_{\rm {bb}}$=$2195\pm 1263$, $kT_{\rm {bb}}$=$2.52\pm 0.18$ keV, and $l_{\rm{nth}}/l_{\rm {h}}$=$0.38\pm 0.19$. The fitting parameters for the two other components are similar to those obtained in the fitting of (a). The parameter errors are inferred at the 90\% confidence level $(\Delta\chi^2=2.7)$.} \label{fig:eufs_2-200keV}
\end{figure}

As shown in Figure~\ref{fig:hard_tail} or Figure~\ref{Fig:fitting_HE_ScoX-1}, the PL component in the HE spectrum of Sco X-1 dominates the flux in the energy band of $\sim$50-200 keV. Such nonthermal component cannot be explained with the hot-corona mechanism, because, as analyzed above, there does not exist a hot corona in Sco X-1. Such high-energy emission cannot be originated from a possible warm corona in this source too, because the warm corona is only with the temperature of several keV. Therefore, other mechanisms responsible for the hard X-ray tail in Sco X-1 should be considered. In XRBs, the accretion process is characterized by the converging inflow onto the compact star and the Comptonization connected with the energetic electrons in the accreting inflow cannot be neglected. \cite{Titarchuk1997} and \cite{Laurent1999} investigated such Comptonization in BHXBs and they called it BMC, i.e. bulk-motion Comptonization. The BMC model includes three parameters, i.e. the temperature of the thermal source for providing seed photons ($kT_ {\rm {bb}}$), the energy spectral index $\alpha$ ($\Gamma=\alpha+1$, $\Gamma$: photon index of the PL component in the spectrum), and $logA$ ($A$: a parameter to describe Comptonization), as well as a normalization ($N$). According to the extent that the seed photons are Comptonized, they are divided into two portions. Some seed photons are weakly Comptonized  and they escape with the way approximating BB emission, which contribute 1/(1+A) of the total luminosity. However,  others are strongly Comptonized and the Comptonzation emission of these seed photons takes A/(1+A) of the total luminosity. The BMC model has been applied to BHXBs  \citep[e.g.,][etc.]{Titarchuk2009,Titarchuk2010,Shrader2010}. It was also used to explain the X-ray nonthermal emission in NS-LMXB Z sources \citep[e.g.,][]{Farinelli2007,Ding2015,Ding2021}. In this work, we assume that the hard X-ray PL tail in Sco X-1 could be resulted from the upscattering of the photons of the BB emission of the NS surface by the free-falling electrons onto the NS in the converging inflow. Meanwhile, taking into account the the iron line and the disk emission in the spectrum, we form a tri-component model of BMC+LINE+CPL to fit the LE+ME+HE spectrum in $\sim$2-200 keV. When fitting, in order to account for the flux calibration differences among the LE, ME, and HE instruments, the model is multiplied by a constant, which is fixed for the LE spectral normalization, but free for the ME and HE spectral normalizations. At the same time, the interstellar photoelectric absorption is considered, with  $N_{\rm {H}}$ being fixed at $\rm {0.3\times 10^{22}\ atom\ cm^{-2}}$ \citep{Church2012}. Panel (a) of Figure~\ref{fig:eufs_2-200keV} demonstrates such a fitting. The value of $\chi^2_{\rm {\nu }}$, as well as the residual distribution, shows that this fitting is statistically acceptable well. It is noted that the seed photon temperature is $\sim$3 keV, indicating that the seed photons for BMC do come from the NS surface emission rather than the disk emission. The value of $logA$ is -1.48, meaning that the BMC emission only occupies $\sim$3\% of the total luminosity, and, therefore, the observed luminosity mainly comes from the direct BB emission of the NS. Actually, comparing the individual components in panel (a) of Figure~\ref{fig:eufs_2-200keV} with those in Figure~\ref{Fig:fitting_LE+ME_spectrum}, one can see that the segment of BMC (dash line) in $\sim$2-35 keV in  panel (a) of Figure~\ref{fig:eufs_2-200keV}, representing the observed direct BB emission of the NS surface, almost happens to correspond the BB component (dash line) in Figure~\ref{Fig:fitting_LE+ME_spectrum}, and, meanwhile, the segment of BMC in $\sim$35-200 keV in  panel (a) of Figure~\ref{fig:eufs_2-200keV}, demonstrating the BMC emission, just is the extension of the BB component of Figure~\ref{Fig:fitting_LE+ME_spectrum} in the high-energy interval. Moreover, the other two components in the two Figures, i.e. the CPL and LINE components, just correspond to each other. In conclusion, the joint fitting of the LE+ME+HE spectrum in $\sim$2-200 keV with the BMC+LINE+CPL model suggests an alternative mechanism for producing the hard X-ray nonthermal emission in Sco X-1, which is that the hard X-ray PL emission in Sco X-1 could be generated from upscattering of the photons of the BB emission of the NS surface by the energetic free-falling electrons in converging inflow onto the NS.

Although the fitting of the broadband spectrum with the BMC+LINE+CPL model proposes a scenario to produce the hard X-ray tail in Sco X-1, yet there exists a potential contradiction in this interpretation. As pointed out by \cite{Zdziarski2001} and \cite{Revnivtsev2014}, if the BMC process does be responsible for the hard X-ray tail in XRBs, the high-energy cutoff lower than $\sim$100-200 keV should be present in the hard X-ray spectrum of theirs \citep{Titarchuk1997}. However, in this work, our analysis suggests that the hard X-ray tail of Sco X-1 shows a PL-like shape without cutoff in the HE spectrum. Therefore, we investigate other possible mechanisms to generate the hard X-ray tail in Sco X-1. It is a natural assumption that through heating mechanisms, two types of electrons are produced in plasma, one of which consists of the thermal electrons with Maxwellian distribution and the other of which comprises of the nonthermal electrons with PL distribution. Correspondingly, two types of Comptonization processes will take place in the plasma. \cite{Coppi1999} investigated the hybrid Comptonization in the thermal/nonthermal plasmas and developed a model (EQPAIR in XSPEC) to describe such Comptonization emission. In the EQPAIR model, the total luminosity of the source is expressed with the dimensionless compactness, defined as
\begin{equation}
l=\frac{L}{R}\frac{\sigma_{\rm  {T}}}{m_{\rm  {e}}c^3}
\end{equation}  
Where, $L$ is the observed source luminosity, $R$ is the characteristic radius of the hard X-ray emission region approximated as a sphere, $\sigma_{\rm {T}}$ is the Thomson cross section, $m_{\rm {e}}$ is the electron mass, and $c$ is the speed of light. The source luminosity ($l$) consists of hard and soft luminosities, i.e. $l=l_{\rm {h}} + l_{\rm {s}}$. The hard luminosity is assumed to be resulted from the hybrid Comptonization, therefore, $l_{\rm {h}} =l_{\rm {nth}}+ l_{\rm {th}}$, where $l_{\rm {nth}}$ and $l_{\rm {th}}$ are the nonthermal and thermal Comptonization luminosities, respectively. According to Kirkhoff$^{'}$s law, $l_{\rm {nth}}$ and $l_{\rm {th}}$ can also be considered as the powers supplied to heat the nonthermal and thermal electrons, respectively. This model includes twenty parameters, plus a normalization. Among the  parameters, there is a critical parameter, i.e.  $l_{\rm {nth}}/l_{\rm {h}}$, which describes the fraction of the total heating power to heat the nonthermal electrons. When the value of this parameter is 1 or 0, the hybrid Comptonization is reduced to a pure nonthermal Comptonization or a pure thermal Comptonization. Another important parameter is $G_{\rm {inj}}$, which describes the distribution of the nonthermal electrons by $\gamma^{-G_{\rm  {inj}}}$, where $\gamma$ is the Lorentz factor of the nonthermal electrons,  taking the value from $\gamma_{\rm {min}}$ to $\gamma_{\rm {max}}$. In the version 12.11.1 of XSPEC, with which we perform analysis in this work, the ranges of $\gamma_{\rm {min}}$ and $\gamma_{\rm {max}}$ are set to be 1.2-1000 and 5-10000, respectively. This model has been applied to interpret the hard X-ray emission of BHXBs \citep[e.g.,][etc.]{Gierlinski2003,Caballero-Garcia2009,Kalemci2016,Zdziarski2021}. It was also used to explain the hard X-ray tail of Z sources \citep{Farinelli2005,DAi2007}. In this work, we replace  BMC with EQPAIR in the BMC+LINE+CPL model to fit the broadband spectrum of Sco X-1 in $\sim$2-200 keV, in which a hard X-ray tail is present. The fitting is statistically acceptable well and a fitting is shown by Panel (b) of Figure~\ref{fig:eufs_2-200keV}. It is noted that the BB temperature ($kT_{\rm {bb}}$) is around 2.5 keV, indicating that the power supplied to heat plasma particles comes from the NS or the boundary layer between the NS and the disk, rather than the disk, which is consistent with the result of \cite{Farinelli2005} or \citep{DAi2007}. While, without doubt, in BHXBs, such power is provided by the disk, because of the fact that the seed photon temperature $kT_{\rm {bb}}$ for the hybrid Comptonization in BHXBs is lower than $\sim$1 keV \citep[e.g.,][]{Zdziarski2001,Caballero-Garcia2009,Kalemci2016,Cangemi2021}. The value of $l_{\rm {nth}}/l_{\rm {h}}$ is about 0.4, meaning that after being powered, the plasma becomes a hybrid and the Comptonization emission of the thermal electrons predominates over that of the nonthermal electrons. This fitting of ours suggests that the hybrid Comptonization taking place around the NS or in the boundary layer between the NS and the disk could be another alternative mechanism to result in the hard X-ray tail of sco X-1, as proposed by \cite{DAi2007}.

\subsubsection{The peculiar hard X-ray tail in the FB of Sco X-1}

Usually, a hard X-ray tail of Z sources is fitted by a PL with a positive photon index, but, in this work we detect a hard tail with a negative photon index in the FB (see Table~\ref{tab:fitting_hard_tail}). This peculiar hard tail of the same source was also detected previously in the same HID branch with the observations of {\it RXTE} \citep{DAmico2001,Ding2021}. Neither the bulk-motion Comptonization mechanism nor the hybrid Comptonization mechanism can be used to explain  such peculiar hard tail, because the photon index $\Gamma$ in the BMC model or the EQPAIR model is set to be a positive parameter in XSPEC. Actually, in our practice, the broadband spectrum (2-200 keV) with this unusual hard tail cannot be fitted by the BMC+LINE+CPL model or the EQPAIR+LINE+CPL model. The origin of this peculiar hard tail is worth thinking deeply. Firstly, it is found in the flaring branch. Generally, the FB of Z sources is considered to be resulted from the unstable nuclear burning on the NS surface. Therefore, is this unusual hard tail connected with the unstable nuclear burning on the NS surface? Secondly, as the argument of \cite{Ding2021}, is it linked to the jet in NS-LMXBs? Thirdly, it seems that among the Z sources, the unusual hard tail is only detected in Sco X-1, so, does it has to do with the source?

\subsection{Distinguishing between BHXBs and NS-LMXBs by hard X-ray spectra}

This work of ours, as well as some works of others, shows that the hard X-ray spectrum of BHXBs differentiates much from that of the high-luminosity NS-LMXBs, i.e. the Z sources. On one hand, the hard X-ray spectrum of BHXBs reveals high-energy cutoff in most cases, but the cutoff phenomenon does not appear in the hard X-ray spectrum of Z sources. On the other hand, the hard X-ray spectrum of BHXBs must be fitted with the one-component mode of CPL in most cases. Occasionally, it can be fitted by a PL or a CPL, in which the the fitting is statistically better with the CPL than with the PL. Nevertheless, the spectral model of the hard X-ray spectrum of Z sources should consists of two components. One could be a thermal component related to the NS emission and the other must be the nonthermal component described with the PL. Interestingly, the spectrum of the low-luminosity NS-LMXBs, i.e. the Atoll sources, sometimes behaves similarly to that of BHXBs \citep{Rodi2016}.

As demonstrated in the Figure 13 of \cite{Barret2000}, the hard X-ray luminosities of BHXBs generally exceed those of low-luminosity NS-LMXBs, i.e. Atoll sources. Here, we compare the hard X-ray luminosities between Cyg X-1 and Z-source Sco X-1. Assuming the distances of Cyg X-1 and Sco X-1 to be 1.86 kpc \citep{Reid2011} and 2.8 kpc \citep{Bradshaw1997}, respectively, we estimate the hard X-ray luminosities of the two sources. The inferred luminosities in 20-200 keV of Cyg X-1 and Sco X-1 span a range of $\sim$(8.5-10.1)$\times10^{36}$ ergs s$^{-1}$ and a range of $\sim$(6.5-12.9)$\times10^{36}$ ergs s$^{-1}$, respectively. Therefore, it seems that the hard X-ray luminosity of the high-luminosity Z sources could be comparable to that of BHXBs, which might be owing to the NS emission of Z sources. As analyzed above, in Z sources, the thermal emission of the NS contribute a portion of low-energy hard X-ray flux . So, reducing the hard X-ray interval, we make further comparison. The derived luminosities in 30-200 keV of Cyg X-1 and Sco X-1 are  $\sim$(6.9-8.4)$\times10^{36}$ ergs s$^{-1}$ and $\sim$(1.0-2.6)$\times10^{36}$ ergs s$^{-1}$, respectively. In this range, the hard X-ray luminosity is obviously larger of the former than of the latter. Furthermore, the higher the hard X-ray energy band, the larger the luminosity difference. These spectral characteristics could provide a way to distinguish the two types of XRBs. 

\section{CONCLUSION}

In this work, with the observations of {\it Insight}-HXMT for Sco X-1, we search for the PL component in its hard X-ray spectrum and investigate the origin of this nonthermal emission. We detect eight hard X-ray tails of this Z source. The detected hard tails are distributed throughout the Z-track on the HID. On the HID, along the Z-track, the hard tail becomes hard and fades away from HB, via NB, to FB. Studying the hard X-ray spectrum of Cyg X-1 and comparing it with that of Sco X-1, it is concluded that the former shows high-energy cutoff, but the latter does not, indicating that there might exist a hot corona in Cyg X-1, but there is without a hot corona in Sco X-1. However, the spectral analysis in $\sim$2-35 keV implies a warm corona in Sco X-1. Fitting the broadband spectrum of Sco X-1 suggests two alternative mechanisms responsible for the hard X-ray tails of Sco X-1. One is that the hard tail could be resulted from the Comptonization of the photons of the NS surface emission by the energetic free-falling electrons onto the NS. The other possibility is that the hard tail might be generated through the hybrid Comptonization of the NS surface emission photons by the thermal as well as nonthermal electrons in the plasmas around the NS or in the boundary layer between the NS and the disk. But, any of the two mechanisms cannot be responsible for the hard tail in the FB. Several possibilities for the production of the peculiar hard tail of the FB are argued, including the unstable nuclear burning on the NS surface, linking to the jet, and connecting with the source. 

\acknowledgments

This research has made use of data from the {\it Insight}-HXMT mission, the first X-ray satellite of China, and the software provided by the {\it Insight}-HXMT team.

\clearpage

\begin{table}
\vspace{3.5cm}
\scriptsize
\caption{Observations of hard X-ray tail detections in Sco X-1}\label{tab:fitting_hard_tail}
\vspace{0.8mm}
\begin{tabular}{lcccccccccccc}
\hline\hline
 & & & & & \multicolumn{2}{c}{BREMSS} & & \multicolumn{2}{c}{PL} & & \\
     \cline{6-7} \cline{9-10}
 OBSID & Dates UT & Z Position & HE Good Time (s) & & $\rm {kT\ (keV)}$ & $^{b}$Flux & & $\Gamma$ & $^{c}$Flux & & $\rm {\chi^2~(dof)}$  & $^{d}$$\rm {F-test}$  \\
\hline
P010132800801  & 2018 Jun. 24 & HB  & $^{a}$1374 & & $4.98_{-0.32}^{+0.31}$ & $1.24_{-0.11}^{+0.11}$ & & $2.26_{-0.38}^{+0.38}$  & $1.09_{-0.11}^{+0.11}$ & & 56.65(52)  & $1.11\times10^{-19}$  \\
P010132801001  & 2018 Aug. 16 & HB &     766         & & $4.96_{-0.26}^{+0.24}$ & $1.35_{-0.15}^{+0.14}$ & & $2.44_{-0.47}^{+0.50}$  & $1.43_{-0.18}^{+0.18}$ & & 26.94(39)  & $5.59\times10^{-17}$  \\
P010132801002  & 2018 Aug. 16 & HB &     2033       & & $5.00_{-0.16}^{+0.15}$ & $1.30_{-0.09}^{+0.08}$ & & $1.98_{-0.36}^{+0.37}$  & $1.19_{-0.12}^{+0.13}$ & & 54.65(54)  & $5.72\times10^{-21}$  \\
P010132801003  & 2018 Aug. 16 & HB &     3126       & & $4.57_{-0.17}^{+0.16}$ & $1.00_{-0.13}^{+0.10}$ & & $3.38_{-0.56}^{+0.71}$  & $0.77_{-0.09}^{+0.09}$ & & 20.40(29)  & $1.87\times10^{-14}$  \\
\hline
P010132800802  & 2018 Jun. 24 & NB &     1859        & & $4.43_{-0.17}^{+0.16}$ & $0.72_{-0.06}^{+0.06}$ & & $1.98_{-0.40}^{+0.43}$  & $0.81_{-0.11}^{+0.12}$ & & 22.37(37)  & $3.71\times10^{-17}$  \\
P010132800804  & 2018 Jun. 24 & NB &     601          & & $4.68_{-0.26}^{+0.25}$ & $0.85_{-0.09}^{+0.09}$ & & $1.12_{-0.58}^{+0.57}$  & $1.27_{-0.26}^{+0.28}$ & & 18.62(33)  & $1.44\times10^{-12}$  \\
P010132800805  & 2018 Jun. 25 & NB &     3846        & & $4.67_{-0.13}^{+0.13}$ & $0.84_{-0.05}^{+0.05}$ & & $1.83_{-0.39}^{+0.39}$  & $0.76_{-0.09}^{+0.09}$ & & 97.84(61)  & $6.06\times10^{-17}$  \\
\hline
P010132800401  & 2018 Apr. 3  & FB &     3687        & & $4.46_{-0.10}^{+0.10}$ & $0.57_{-0.02}^{+0.02}$ & & $-1.13_{-0.61}^{+0.58}$ & $0.51_{-0.12}^{+0.11}$ & & 34.15(29)  & $9.21\times10^{-8}$   \\
\hline
\end{tabular}
\tablecomments{Fitting is performed with the BREMSS+PL model in $\sim$30-200 keV. The parameter
errors are derived at the 90\% confidence level ($\Delta\chi^2=2.7$).}
\tablenotetext{a}{The observation of this OBSID spans in the HB and the NB. This interval  is the exposure time in the HB.}
\tablenotetext{b}{Flux in the 30-60 keV range, in units of $10^{-9}$ ergs cm$^{-1}$ s$^{-1}$.}
\tablenotetext{c}{Flux in the 30-200 keV range, in units of $10^{-9}$ ergs cm$^{-1}$ s$^{-1}$.}
\tablenotetext{d}{$F$-test probability for adding the PL component in the spectral model.}
\end{table}

\clearpage

\begin{table}
\vspace{3.5cm}
\tiny
\caption{Fitting result of the HE spectrum of Cyg X-1}\label{tab:HE_Cyg X-1}
\vspace{0.8mm}
\begin{tabular}{lcccccccccccc}
\hline\hline
& & & & \multicolumn{4}{c}{CPL} & & \multicolumn{4}{c}{COMPTT}  \\
                  \cline{5-8}                                  \cline{10-13}
OBSID & Dates UT & HE Good Time (s) &  & $\Gamma$ & $\rm {E_{cut}\ (keV)}$ & $^{a}$Flux & $\rm {  \chi^2_{\nu}~(dof)}$ & & $\rm {kT_{e}\ (keV)} $ & $\rm {\tau}$ & N & $\rm {  \chi^2_{\nu}~(dof)}$  \\
\hline
P020101216002 & 2019 Jul. 11    &     2838  &  & $1.53_{-0.06}^{+0.06}$ & $148_{-18}^{+24}$  & $17.33_{-0.18}^{+0.18}$ & 0.93(88)    &  &  $45.4_{-4.0}^{+6.5}$     & $1.16_{-0.16}^{+0.13}$  & $0.15_{-0.02}^{+0.01}$      & 0.90(88)      \\
P020101216003 & 2019 Jul. 11    &     1536  &  & $1.50_{-0.07}^{+0.07}$ & $141_{-20}^{+27}$  & $18.49_{-0.22}^{+0.22}$ & 1.18(85)    &  &  $43.7_{-4.2}^{+7.1}$     & $1.21_{-0.18}^{+0.14}$  & $0.17_{-0.02}^{+0.01}$      & 1.18(85)      \\
P020101216401 & 2019 Jul. 20    &     2401  &  & $1.51_{-0.06}^{+0.06}$ & $198_{-29}^{+40}$  & $18.76_{-0.18}^{+0.19}$ & 1.31(97)    &  &  $85.2_{-28.7}^{+46.4}$  & $0.68_{-0.29}^{+0.38}$  & $0.07_{-0.02}^{+0.03}$      & 1.34(97)      \\
P020101216501 & 2019 Jul. 21    &     4072  &  & $1.46_{-0.05}^{+0.05}$ & $162_{-18}^{+22}$  & $18.88_{-0.16}^{+0.16}$ & 1.04(96)    &  &  $47.1_{-3.5}^{+5.1}$     & $1.25_{-0.12}^{+0.10}$  & $0.12_{-0.01}^{+0.01}$      & 1.05(96)      \\
P020101216601 & 2019 Jul. 22    &     4472  &  & $1.49_{-0.04}^{+0.04}$ & $173_{-16}^{+20}$  & $21.02_{-0.14}^{+0.14}$ & 1.29(106)  &  &  $59.6_{-8.5}^{+37.2}$    & $0.98_{-0.42}^{+0.17}$  & $0.12_{-0.01}^{+0.01}$      & 1.45(106)    \\
P020101217201 & 2019 Aug. 03  &     3181  &  & $1.58_{-0.06}^{+0.06}$ & $187_{-26}^{+35}$  & $17.33_{-0.16}^{+0.16}$ & 1.21(96)    &  &  $72.5_{-20.3}^{+42.1}$  & $0.72_{-0.32}^{+0.30}$  & $0.09_{-0.03}^{+0.03}$      & 1.23(96)      \\
\hline
\end{tabular}
\tablecomments{The spectrum is fitted with the CPL model and the COMPTT model, respectively. Fitting is performed in 30-220 keV. When fitting with the COMPTT model, the soft photon (Wien) temperature is fixed at the default value, i.e. 0.1 keV. The parameter errors are inferred at the 90\% confidence level ($\Delta\chi^2=2.7$).}
\tablenotetext{a}{Flux in the 30-220 keV range, in units of $10^{-9}$ ergs cm$^{-1}$ s$^{-1}$.}
\end{table}

\clearpage

\clearpage

\clearpage

\clearpage

\clearpage

\clearpage

\clearpage

\begin{thebibliography}{}

 \bibitem[Barret et al.(2000)]{Barret2000}Barret, D., Olive, J. F., Boirin, L., et al. 2000, \apj, 533, 329 
  
\bibitem[Blandford \& Payne(1982)]{Blandford1982}Blandford, R. D., \& Payne, D. G. 1982, \mnras, 199, 883 
  
\bibitem[Blandford \& Znajek(1977)]{Blandford1977}Blandford, R. D., \& Znajek, R. L. 1977, \mnras, 179, 433 

\bibitem[Bradshaw et al.(1997)]{Bradshaw1997}Bradshaw, C. F., Fomalont, E. B., \& Geldzahler, B. J. 1997, \apjL, 484, L55
  
\bibitem[Caballero-Garc{\'i}a et al.(2009)]{Caballero-Garcia2009}Caballero-Garc{\'i}a, M. D., Miller, J. M., Trigo, M. D., et al. 2009, \apj, 692, 1339

\bibitem[Cangemi et al.(2021)]{Cangemi2021}Cangemi, F., Beuchert, T., Siegert, T., et al. 2021, \aap, 650, A93
  
\bibitem[Chen et al.(2018)]{Chen2018}Chen, Y. P., Zhang, S., Qu, J. L., et al. 2018, \apjl, 864, L30

\bibitem[Church \&  Ba{\l}uci{\'n}ska-Church(1995)]{Church1995}Church, M. J., \& Ba{\l}uci{\'n}ska-Church, M. 1995, \aap, 300, 441

\bibitem[Church \& Ba{\l}uci{\'n}ska-Church(2004)]{Church2004}Church, M. J., \& Ba{\l}uci{\'n}ska-Church, M. 2004, \mnras, 348, 955
  
\bibitem[Church et al.(2012)]{Church2012}Church, M. J., Gibiec, A., Ba{\l}uci{\'n}ska-Church, M., \& Jackson, N. K. 2012, \aap, 546, A35

\bibitem[Coppi(1999)]{Coppi1999}Coppi, P. S. 1999, in ASP Conf. Ser. 161, High Energy Processes in Accreting Black Holes, ed. J. Poutanen \& R. Svensson (San Francisco, CA: ASP), 375

\bibitem[D'Ai et al.(2007)]{DAi2007}D'A{\'i}, A., {\.Z}ycki, P., Di Salvo, T., Iaria, R., Lavagetto, G., \& Robba, N. R. 2007, \apj, 667, 411
  
\bibitem[D'Amico et al.(2001)]{DAmico2001}D'Amico, F., Heindl, W. A., Rothschild, R. E., \& Gruber, D. E. 2001, \apjl, 547, L147

\bibitem[Ding et al.(2021)]{Ding2021}Ding, G. Q., Chen, T. T., \& Qu, J. L. 2021, \mnras, 500, 772

\bibitem[Ding \& Huang(2015)]{Ding2015}Ding, G. Q., \& Huang, C. P. 2015, JApA, 36, 335

\bibitem[Ding et al.(2003)]{Ding2003}Ding, G. Q., Qu, J. L., \& Li, T. P. 2003, \apjl, 596, L219

\bibitem[Ding et al.(2006a)]{Ding2006a}Ding, G. Q., Qu, J. L., \& Li, T. P. 2006a, \aj, 131, 1693
  
\bibitem[Ding et al.(2006b)]{Ding2006b}Ding, G. Q., Zhang, S. N., Li, T. P., \& Qu, J. L. 2006b, \apj, 645, 576

\bibitem[Ding et al.(2011)]{Ding2011}Ding, G. Q., Zhang, S. N., Wang, N., Qu, J. L., \& Yan, S. P. 2011, \aj, 142, 34
  
\bibitem[DiSalvo et al.(2002)]{DiSalvo2002}Di Salvo, T., Farinelli, R., Burderi, L., et al. 2002, \aap, 386, 535

\bibitem[Di Salvo et al.(2006)]{DiSalvo2006}Di Salvo, T., Goldoni, P., Stella, L., et al. 2006, \apj, 649, L91

\bibitem[Di Salvo et al.(2001)]{DiSalvo2001}Di Salvo, T., Robba, N. R., Iaria, R., et al. 2001, \apj, 554, 49
  
\bibitem[Di Salvo et al.(2000)]{DiSalvo2000}Di Salvo, T., Stella, L., Robba, N. R., et al. 2000, \apjL, 544, L119
  
\bibitem[Dove et al.(1998)]{Dove1998}Dove, J. B., Wilms, J., Nowak, M. A., Vaughan, B. A., \& Begelman, M. C. 1998, \mnras, 298, 729 

\bibitem[Farinelli et al.(2007)]{Farinelli2007}Farinelli, R. Titarchuk, L., \& Frontera, F. 2007, \apj, 662, 1167

\bibitem[Farinelli et al.(2005)]{Farinelli2005}Farinelli, R., Frontera, F., Zdziarski, A. A., et al. 2005, \aap, 434, 25

\bibitem[Fiocchi et al.(2006)]{Fiocchi2006}Fiocchi, M., Bazzano, A.,, Ubertini, P., \& Jean, P. 2006, \apj, 651, 416

\bibitem[Galeev et al.(1979)]{Galeev1979}Galeev, A. A., Rosner, R., \& Vaiana, G. S. 1979, \apj, 229, 318
  
\bibitem[Ge et al.(2020)]{Ge2020}Ge, M. Y., Ji, L., Zhang, S. N., et al. 2020, \apjL, 899, L19

\bibitem[Ghosh \& Lamb (1979)]{Ghosh1979}Ghosh, P., \& Lamb, F. K. 1979, \apj, 234, 296 

\bibitem[Gierli{\'n}ski \& Done(2003)]{Gierlinski2003}Gierli{\'n}ski, M., \& Done, C. 2003, \mnras, 342, 1083 
  
\bibitem[Huang et al.(2018)]{Huang2018}Huang, Y., Qu, J. L., Zhang, S. N., et al. 2018, \apj, 866, 122

\bibitem[Lamb et al.(1973)]{Lamb1973}Lamb, F. K., Pethick, C. J., \& Pines, D. 1973, \apj, 184, 271

\bibitem[Lamb \& Sanford(1979)]{Lamb1979}Lamb, P., \& Sanford, P. W. 1979, \mnras, 188, 555 
  
\bibitem[Iaria et al.(2001)]{Iaria2001}Iaria, R., Burderi, L., Di Salvo, T., La Barbera, A., \& Robba, N. R. 2001, \apj, 547, 412

\bibitem[Iaria et al.(2002)]{Iaria2002}Iaria, R., Di Salvo, T., Robba, N. R., \& Burderi, L. 2002, \apj, 567, 503

\bibitem[Kalemci et al.(2016)]{Kalemci2016}Kalemci, E., Begelman, M. C., Maccarone, T. J., et al. 2016, \mnras, 463, 615
  
\bibitem[Laurent \& Titarchuk(1999)]{Laurent1999}Laurent, P., \& Titarchuk, L. 1999, \apj, 511, 289
  
\bibitem[Lavagetto et al.(2004)]{Lavagetto2004}Lavagetto, G., Iaria, R., di Salvo, T., et al. 2004, NuPhS, 132, 616
  
\bibitem[Liu et al.(2011)]{Liu2011}Liu, B. F., Done, C., \& Taam, R.  E. 2011, \apj, 726, 10 

\bibitem[Liu et al.(2003)]{Liu2003}Liu, B. F., Mineshige, S., \& Ohsuga, K. 2003, \apj, 587, 571
  
\bibitem[Liu et al.(2002)]{Liu2002}Liu, B. F., Mineshige, S., \& Shibata, K. 2002, \apjL, 572, L173 
  
\bibitem[Li et al.(2020)]{Li2020}Li, X. B., Li, X. F., Tan, Y., et al. 2020, JHEAp, 27, 64
  
\bibitem[Li et al.(2018)]{Li2018}Li, X. B., Song L. M., Li X.F., et al. 2018, Proc. SPIE,
10699, 1069969

\bibitem[Li et al.(2019)]{Li2019}Li, X. F., Li, C. Z., Chang, Z., et al. 2019, JHEAp, 24, 6

\bibitem[Ma et al.(2021)]{Ma2021}Ma, X., Tao, L.,  Zhang, S. N., et al. 2021, Nature Astronomy, 5, 94
  
\bibitem[Mitsuda et al.(1984)]{Mitsuda1984}Mitsuda, K., Inoue, H., Koyama, K., et al. 1984, \pasj, 36, 741

\bibitem[Mitsuda et al.(1989)]{Mitsuda1989}Mitsuda, K., Inoue, H., Nakamura, N., \& Tanaka, Y. 1989, \pasj, 41, 97
  
\bibitem[Montanari et al.(2009)]{Montanari2009}Montanari, E., Titarchuk, L., \& Frontera, F. 2009, \apj, 692, 1597 

\bibitem[Motta et al.(2021)]{Motta2021}Motta, S. E., Rodriguez, J., Jourdain, E., et al. 2021, New Astronomy Reviews (NewAR), 93, 101618

\bibitem[Mukerjee et al.(2020)]{Mukerjee2020}Mukerjee, K., Antia, H. M., \& Katoch, T. 2020, \apj, 897, 73 

\bibitem[Nishimura et al.(1986)]{Nishimura1986}Nishimura, J., Mitsuda, K., \& Itoh, M. 1986, \pasj, 38, 819
  
\bibitem[Paizis et al.(2005)]{Paizis2005}Paizis, A., Ebisawa, K., Tikkanen, T., et al. 2005, \aap, 443, 599

\bibitem[Petrucci et al.(2001)]{Petrucci2001}Petrucci, P. O., Haardt, F., Maraschi, L., et al. 2001, \apj, 556, 716  
  
\bibitem[Piraino et al.(1999)]{Piraino1999}Piraino, S., Santangelo, A., Ford, E. C., \& Kaaret, P. 1999, \aap, 349, L77  
  
\bibitem[Qiao \& Liu(2012)]{Qiao2012}Qiao, E., \& Liu, B. F. 2012, \apj,  744, 145

\bibitem[Raichur et al.(2011)]{Raichur2011}Raichur, H., Misra, R., \& Dewangan, G. 2011, \mnras, 416, 637

\bibitem[Remillard \& McClintock(2006)]{Remillard2006}Remillard, R. A., \& McClintock, J. E. 2006, \araa, 44, 49

\bibitem[Reid et al.(2011)]{Reid2011}Reid, M. J., McClintock, J. E., Narayan, R., et al. 2011, \apj, 742, 83
  
\bibitem[Revnivtsev et al.(2014)]{Revnivtsev2014}Revnivtsev, M. G., Tsygankov, S. S., Churazov, E. M., \& Krivonos, R. A. 2014, \mnras, 445, 1205   

\bibitem[Rodi et al.(2016)]{Rodi2016}Rodi, J., Jourdain, E., \& Roques, J. P. 2016, \apj, 817, 101
  
\bibitem[Shrader et al.(2010)]{Shrader2010}Shrader, C. R., Titarchuk, L., \& Shaposhnikov, N. 2010, \apj, 718, 488 
  
\bibitem[Sunyaev \& Titarchuk(1980)]{Sunyaev1980}Sunyaev, R. A., \& Titarchuk, L. G. 1980, \aap, 86, 121  
  
\bibitem[Revnivtsev et al.(2014)]{Revnivtsev2014}Revnivtsev, M. G., Tsygankov, S. S., Churazov, E. M., \& Krivonos, R. A. 2014, \mnras, 445, 1205
  
\bibitem[Tarana et al.(2007)]{Tarana2007}Tarana, A., Bazzano, A., Ubertini, P., \& Zdziarski, A. A. 2007, \apj, 654, 494 

\bibitem[Tarana et al.(2011)]{Tarana2011}Tarana, A., Belloni, T., Bazzano, A., M{\'e}ndez, M., \& Ubertini, P. 2011, \mnras, 416, 873

\bibitem[Titarchuk (1994)]{Titarchuk1994}Titarchuk, L. 1994, \apj, 434, 570

\bibitem[Titarchuk et al.(1997)]{Titarchuk1997}Titarchuk, L., Mastichiadis, A., \& Kylafis, N. D. 1997, \apj, 487, 837 
  
\bibitem[Titarchuk \& Seifina(2009)]{Titarchuk2009}Titarchuk, L., \& Seifina, E. 2009, \apj, 706, 1463 
    
\bibitem[Titarchuk \& Shaposhnikov(2010)]{Titarchuk2010}Titarchuk, L., \& Shaposhnikov, N. 2010, \apj, 724, 1147 

\bibitem[Varun et al.(2019)]{Varun2019}Varun, Maitra, C., Pragati, P., Harsha, R., \& Biswajit, P. 2019, \mnras, 484, L1

\bibitem[White et al.(1986)]{White1986}White, N. E., Peacock, A., Hasinger, G., et al. 1986, \mnras, 218, 129
  
\bibitem[White et al.(1985)]{White1985}White, N. E., Peacock, A., \& Taylor, B. G. 1985, \apj, 296, 475
  
\bibitem[Yan et al.(2020)]{Yan2020}Yan, Z., Xie, Fu-Guo, \& Zhang, W. 2020, \apjL, 889, L18
  
\bibitem[Yao et al.(2005)]{Yao2005}Yao, Y.,  Zhang, S. N.,  Zhang, X., Feng, Y., \& Robinson, C.  R. 2005, \apj, 619, 446

\bibitem[You et al.(2021)]{You2021}You, B., Tou, Y., Li, C., et al. 2021, Nature Communications, 12, 1025  

\bibitem[Zdziarski et al.(2001)]{Zdziarski2001}Zdziarski, A. A., Grove, J. E., Poutanen, J., Rao, A. R., \& Vadawale, S. V. 2001, \apjL, 554, L45 

\bibitem[Zdziarski et al.(2021)]{Zdziarski2021}Zdziarski, A. A., Jourdain, E., Lubi{\'n}ski, P., et al. 2021, \apjL, 914, L5
  
\bibitem[Zhang et al.(2014)]{Zhang2014}Zhang, S., Lu, F. J., Zhang, S. N., \& Li, T. P. 2014, Proc. SPIE, 9144, 914421
  
\bibitem[Zhang et al.(2000)]{Zhang2000}Zhang, S. N., Cui, W.,  Chen, W., et al. 2000,  \sci, 287, 1239

\bibitem[Zhang et al.(2020)]{Zhang2020}Zhang, S. N., Li, T. P., Lu, F. J., et al. 2020, SCPMA, 63, 249502
  
\end{thebibliography}
\end{document}